\documentclass[aps, prd, twocolumn, amsmath, floats,floatfix, superscriptaddress, nofootinbib,
showpacs]{revtex4}

\usepackage{amssymb}
\usepackage{amsmath}
\usepackage{verbatim}
\usepackage{mathrsfs}
\usepackage{amsfonts}
\usepackage{latexsym}
\usepackage{epsfig}
\usepackage{color}
\usepackage{graphicx,subfigure}
\usepackage{units}
\usepackage{slashbox}
\usepackage{envmath}
\usepackage{natbib}

\begin{document}


\definecolor{orange}{rgb}{0.9,0.45,0}

\newcommand{\re}{\mbox{Re}}
\newcommand{\im}{\mbox{Im}}
\newcommand{\tf}[1]{\textcolor{red}{TF: #1}}
\newcommand{\jc}[1]{\textcolor{blue}{JC: #1}}
\newcommand{\nsg}[1]{\textcolor{cyan}{N: #1}}
\newcommand{\ch}[1]{\textcolor{green}{CH: #1}}

\def\CovDev{D}
\def\Res{{\mathcal R}}
\def\Gammaflat{\hat \Gamma}
\def\metricflat{\hat \gamma}
\def\Dflat{\hat {\mathcal D}}
\def\part_n{\partial_\perp}

\def\Lie{\mathcal{L}}
\def\A{\mathcal{X}}
\def\Aphi{\A_{\phi}}
\def\hAphi{\hat{\A}_{\phi}}
\def\E{\mathcal{E}}
\def\Ham{\mathcal{H}}
\def\M{\mathcal{M}}
\def\R{\mathcal{R}}
\def\p{\partial}

\def\hg{\hat{\gamma}}
\def\hA{\hat{A}}
\def\hD{\hat{D}}
\def\hE{\hat{E}}
\def\hR{\hat{R}}
\def\hcA{\hat{\mathcal{A}}}
\def\hDelt{\hat{\triangle}}

\renewcommand{\t}{\times}

\long\def\symbolfootnote[#1]#2{\begingroup%
\def\thefootnote{\fnsymbol{footnote}}\footnote[#1]{#2}\endgroup}


\title{Numerical evolutions of spherical Proca stars} 	
 
\author{Nicolas Sanchis-Gual}
\affiliation{Departamento de
  Astronom\'{\i}a y Astrof\'{\i}sica, Universitat de Val\`encia,
  Dr. Moliner 50, 46100, Burjassot (Val\`encia), Spain}

    \author{Carlos Herdeiro}
\affiliation{Departamento de F\'{\i}sica da Universidade de Aveiro and 
Centre for Research and Development in Mathematics and Applications (CIDMA), 
Campus de Santiago, 
3810-183 Aveiro, Portugal}

 \author{Eugen Radu}
\affiliation{Departamento de F\'{\i}sica da Universidade de Aveiro and 
Centre for Research and Development in Mathematics and Applications (CIDMA), 
Campus de Santiago, 
3810-183 Aveiro, Portugal}

\author{Juan Carlos Degollado} 
\affiliation{
Instituto de Ciencias F\'isicas, Universidad Nacional Aut\'onoma de M\'exico,
Apdo. Postal 48-3, 62251, Cuernavaca, Morelos, M\'exico.}

\author{Jos\'e A. Font}
\affiliation{Departamento de
  Astronom\'{\i}a y Astrof\'{\i}sica, Universitat de Val\`encia,
  Dr. Moliner 50, 46100, Burjassot (Val\`encia), Spain}
\affiliation{Observatori Astron\`omic, Universitat de Val\`encia, C/ Catedr\'atico 
  Jos\'e Beltr\'an 2, 46980, Paterna (Val\`encia), Spain}


\date{February 2017}


\begin{abstract} 
Vector boson stars, or \textit{Proca stars}, have been recently obtained as fully non-linear numerical solutions of the Einstein-(complex)-Proca system~\cite{Brito:2015pxa}. These are self-gravitating, everywhere non-singular, horizonless Bose-Einstein condensates of a massive vector field, which resemble in many ways, but not all, their scalar cousins, the well-known (scalar) \textit{boson stars}. In this paper we report fully-non linear numerical evolutions of Proca stars, focusing on the spherically symmetric case, with the goal of assessing their stability and the end-point of the evolution of the unstable stars.  Previous results from linear perturbation theory indicate the separation between stable and unstable configurations occurs at the solution with maximal ADM mass. Our simulations confirm this result. Evolving numerically unstable solutions, we find, depending on the sign of the binding energy of the solution and on the perturbation, three different outcomes: $(i)$ migration to the stable branch, $(ii)$ total dispersion of the scalar field, or $(iii)$ collapse to a  Schwarzschild black hole. In the latter case, a long lived Proca field remnant -- a \textit{Proca wig} -- composed by quasi-bound states, may be seen outside the horizon after its formation, with a life-time that scales inversely with the Proca mass. We comment on the similarities/differences with the scalar case as well as with neutron stars. 
\end{abstract}




\maketitle

\vspace{0.8cm}

\section{Introduction}
In Minkowski spacetime, the linear, massive, Klein-Gordon equation has a 
general solution given in terms of Fourier modes, corresponding to different 
frequencies. The amplitudes of these modes can be chosen as to yield, at some 
given time, a \textit{lump-like} solution, wherein  all (or almost all) energy 
is localized in a compact spatial region. The different phase velocity of the 
different modes, however, implies this lump will be dispersed by the time 
evolution.

\textit{Soliton-like} solutions of the Klein-Gordon field, corresponding to 
energy lumps that are preserved by the time evolution, require non-linearities, 
that will compensate the field's natural tendency to disperse. These 
non-linearities may be due to the field self-interactions~\cite{Coleman:1985ki}, 
or due to gravity. Examples within the latter context where first discovered 
in~\cite{Kaup:1968zz,Ruffini:1969qy}, as static, spherically symmetric solutions 
of the Einstein--(massive, complex)Klein-Gordon system, and became known as 
(scalar) \textit{boson stars} (SBSs) -- see~\cite{Schunck:2003kk} for a review.

Unlike the (short-lived) flat space lumps described in the first paragraph, 
SBSs correspond to a \textit{single} frequency mode with a very  large 
occupation number. They are sometimes described as relativistic, 
self-gravitating, Bose-Einstein condensates. This single frequency appears as a 
harmonic time dependence for the scalar field. Thus, compatibility with a static 
geometry requires the field to be complex, containing two scalar degrees of 
freedom, both oscillating but with a phase difference of $\pi/2$. Physically, 
these oscillations prevent the collapse of the scalar field into a black hole 
and simultaneously evade no-soliton Derrick~\cite{Derrick:1964ww}/virial type 
theorems. The opposite phases, on the other hand, shut down scalar field 
dissipation towards infinity, which is observed for real fields, for which only 
quasi-stationary configurations exist, albeit extremely long 
lived~\cite{Grandclement:2011wz}, dubbed 
\textit{oscillatons}~\cite{Seidel:1991zh}. For rotating 
SBSs~\cite{Schunck:1996he,Yoshida:1997qf} moreover, the two opposite phases also 
shut down gravitational wave emission.  We remark, that SBSs require a mass term 
but no self-interactions, even though the latter may also be 
included~\cite{Colpi:1986ye,Kleihaus:2007vk,Ryan:1996nk,Grandclement:2014msa,
Herdeiro:2015tia}.

The existence of boson stars as stationary solutions of the Einstein equations 
raises questions about their dynamics (see~\cite{Liebling:2012fv} for a review). 
In particular, one may ask if SBSs: \textit{are stable}, and, even more 
fundamentally, \textit{if they may form dynamically}. For spherically symmetric 
SBSs, the first question was answered at the level of mode stability 
in~\cite{Gleiser:1988ih,Lee:1988av}. Roughly, \textit{fundamental solutions} 
($i.e$ without nodes in the scalar field radial profile) that become too compact 
are unstable against radial perturbations. More precisely, the turnover point, 
stability-wise, corresponds to the solution with the maximum ADM mass for a 
given scalar field mass. This perturbative results were confirmed by fully 
non-linear numerical simulations in~\cite{Seidel:1990jh}, where, moreover, the 
fate of the unstable solutions was reported to be either the formation of a 
black hole, or the migration to the stable branch, regardless of the sign of the 
binding energy. A different possible fate of unstable SBSs with \textit{excess} 
($i.e$ negative binding) energy, is that they disperse entirely, which was 
reported in~\cite{Balakrishna:1997ej} for excited states 
and~\cite{Guzman:2004jw} for fundamental states (see also~\cite{Hawley:2000dt}). 

Concerning the second question above,  the dynamical formation of SBSs was 
first examined, using non-linear numerical simulations, in~\cite{Seidel:1993zk}, 
where it was found that stable SBSs can form starting with generic initial data 
describing an unbound scalar field, after releasing the excess energy via the 
mechanism of \textit{gravitational cooling}. For a discussion of SBSs formation 
in cosmological scenarios see~\cite{Liddle:1993ha}.

\bigskip

\begin{figure}[h!]
\centering
\includegraphics[height=2.45in]{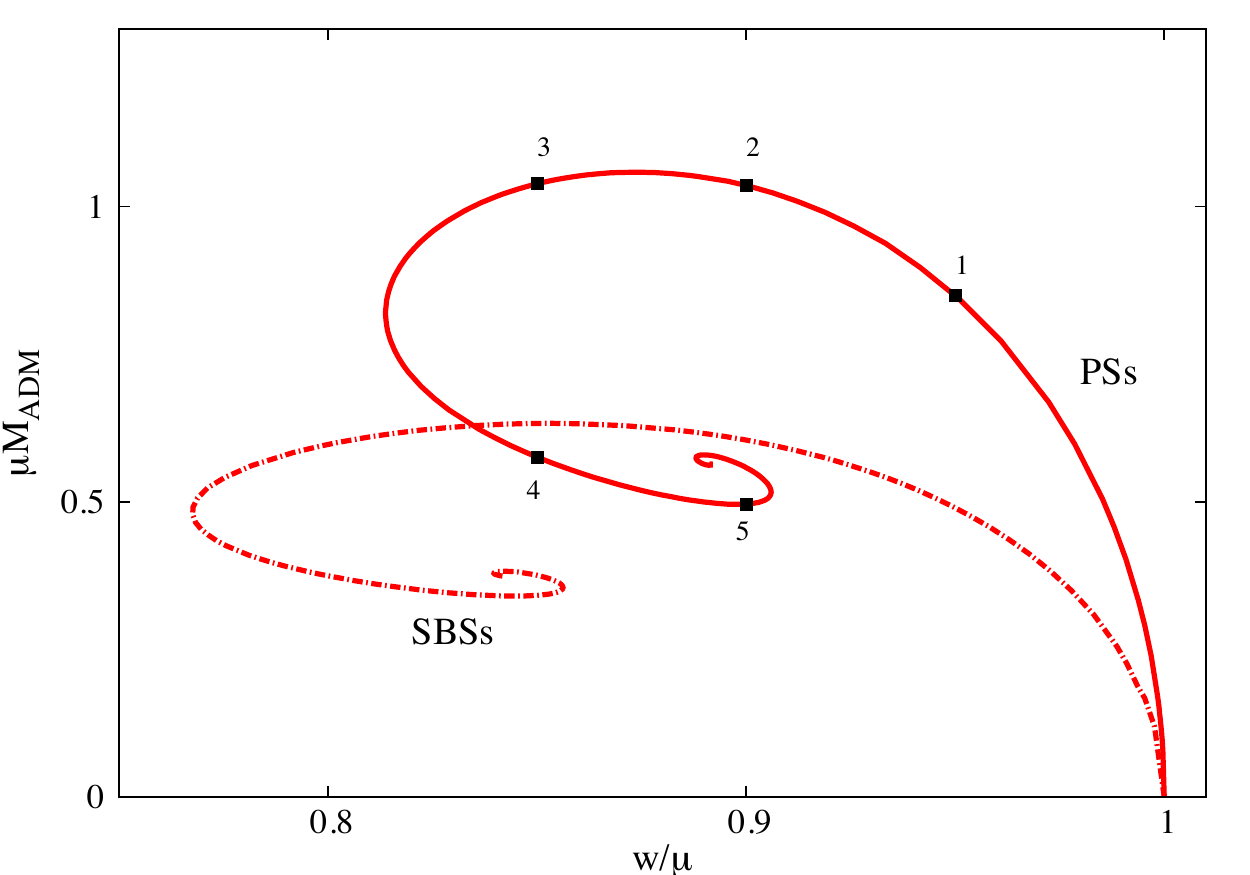}  
\caption{Domain of existence of the spherical (fundamental) PSs (solid line) and spherical (fundamental) SBSs (dashed line) solutions in an ADM mass $vs.$ vector/scalar field frequency
diagram.  Both fields have been assumed to have mass $\mu$. 
The five highlighted points correspond to the configurations to be evolved below.
}
\label{fig1}
\end{figure}

It has long been thought that there could be vector analogues to SBSs, but the 
former have only recently been constructed and are known as~\textit{Proca stars} 
(PSs)~\cite{Brito:2015pxa} - see 
also~\cite{Herdeiro:2016tmi,Garcia:2016ldc,Duarte:2016lig} for generalizations. 
These configurations have been found in the Einstein-(complex)Proca system and, in analogy to their scalar cousins, they correspond to a single frequency 
mode building up a (vector), macroscopic, self-gravitating Bose-Einstein 
condensate. This frequency appears as a harmonic time dependence for the Proca 
potential and the  domain of existence of PSs is very similar to that of SBSs, 
when presented in an ADM mass $vs.$ Proca frequency diagram - Fig.~\ref{fig1} -, 
except that the latter have a lower mass and a wider frequency range.

The stability properties of the spherical PSs solutions also parallel those 
of the spherical SBSs. Analysing radial perturbations, it was shown 
in~\cite{Brito:2015pxa} that there is a stable branch of PSs solutions, 
corresponding in Fig.~\ref{fig1} to the line connecting $w/\mu=1$ to the maximal 
ADM mass, and an unstable branch, corresponding to the remaining part of the 
spiraling curve. The purpose of this paper is to 1) confirm this picture by 
performing fully non-linear time-evolutions, rather than a linear frequency 
domain analysis, and 2) determine the fate of unstable solutions, by evolving a sample of representative cases. 

The evolutions described herein confirm the picture obtained from perturbation 
theory, regarding the existence of a stable branch (connecting the vacuum with 
the solution with maximal ADM mass) and an unstable branch. For the latter, all 
configurations we have evolved with no added perturbation (only the discretization 
error of our numerical code) collapse into a black hole. Introducing a perturbation, 
however, can change the fate of an unstable solution in a way that depends on 
the sign of its binding energy. We illustrate this by a particular type of 
perturbation which makes unstable solutions with positive binding energy migrate 
to the stable branch, whereas unstable solutions with a negative binding energy 
(excess energy) undergo fission; $i.e.$ they disperse entirely. 

For the cases in which a black hole results from the evolution, after apparent horizon formation, we observe the possibility of a long lived remnant Proca field lingering outside the horizon. A Fourier analysis, together with a fit of the Proca remnant, suggest this slowly decaying remnant is composed of quasi-bound states of the Proca field outside a Schwarzschild black hole. This type of \textit{wig} 
around a black hole has been observed for the scalar field case in~\cite{Barranco:2012qs}, 
and its lifetime scales with the inverse of the scalar/Proca field mass.  

\bigskip

This paper is organized as follows. In Section~\ref{sec2} we present the 
covariant Einstein-(complex)Proca model and the equations that will be used for 
the numerical evolutions. In Section~\ref{sec3} we describe the initial data an 
in particular a set of five representative PSs solutions that will be used in 
our numerical evolutions. A brief description of the numerical techniques is 
given in Section~\ref{sec:numerics} and our results are presented in 
Section~\ref{results}. Final remarks are presented in Section~\ref{secconclusions}. 
A brief overview of technical details concerning the static Proca star solutions is given in Appendix~\ref{appendixA} and a succinct assessment of our numerical code is given in Appendix~\ref{appendix}.

\section{Basic equations}
\label{sec2}
We investigate the dynamics of a complex Proca field by solving numerically the Proca equations coupled to the Einstein equations. The system is described by the action $\mathcal{S}=\int d^4x \sqrt{-g}\mathcal{L}$, where the Lagrangian density depends on the Proca potential $\mathcal{A}$, and field strength $\mathcal{F}=d\mathcal{A}$; it is given by:
\begin{equation}
\mathcal{L}=\frac{R}{16\pi 
G_0}-\frac{1}{4}\mathcal{F}_{\alpha\beta}\bar{\mathcal 
{F}}^{\alpha\beta}-\frac{1 } {2}\mu^2\mathcal{A}_\alpha\bar{\mathcal{A}}^\alpha 
\ ,
\label{model}
\end{equation}
where the bar denotes complex conjugation. In the following we briefly describe the main equations used for the numerical implementation, in order to perform time evolutions. 

 \subsection{The BSSN equations}
The evolution of the Einstein-Proca system is performed using the numerical code 
in spherical polar coordinates described in \cite{baumgarte2013numerical}, which 
has been adapted to account for the Proca field (here we set $c=G_{0}=1$). We adopt 
Brown's covariant 
form \cite{brown2009covariant,Alcubierre:2010is} of the BSSN formulation 
\cite{nakamura1987general,shibata1995evolution,baumgarte1998numerical}. The 
conformally related spatial metric $\bar \gamma_{ij}$ is written as
\begin{equation}
\bar \gamma_{ij} = e^{-4 \phi} \gamma_{ij},
\end{equation}
where $\gamma_{ij}$ is the physical spatial metric, and $e^\phi$ a conformal factor. We note that
\begin{equation}
e^{4 \phi} = \left( \bar \gamma / \gamma \right)^{1/3},
\end{equation}
with $\gamma=\det\gamma_{ij}$ and $\bar \gamma= \det\bar \gamma_{ij}$.

We introduce a background connection $\Gammaflat^i_{jk}$ and define
\begin{equation} \label{deltagamma}
\Delta \Gamma^{i}_{jk} = \bar \Gamma^i_{jk} - \Gammaflat^i_{jk}
\end{equation}
which, unlike the two connections themselves, transform as a tensor field. We also define the trace of these variables as
\begin{equation}
\Delta \Gamma^i \equiv \bar \gamma^{jk} \Delta \Gamma^i_{jk}.
\end{equation}
It is not necessary for the background connection to be associated with any metric, but in practice we take it to be the Levi-Civita connection of flat spacetime in spherical coordinates. 

We define the connection vector $\bar \Lambda^i$ as a new set of independent variables that are equal to the $\Delta \Gamma^i$ when the constraint
\begin{equation}
{\mathcal C}^i \equiv \bar \Lambda^i - \Delta \Gamma^i = 0
\end{equation}
holds. Defining $\part_n \equiv \partial_t - {\mathcal L}_\beta$,
where ${\mathcal L}_\beta$ denotes the Lie derivative along the 
shift vector $\beta^i$, we then obtain the following set of evolution equations 
\begin{subequations} \label{evolution}
\begin{eqnarray}
\part_n \bar \gamma_{ij} & = & - \frac{2}{3} \bar \gamma_{ij} \bar D_k \beta^k - 2 \alpha \bar A_{ij} \\
\part_n \bar A_{ij} & = & - \frac{2}{3} \bar A_{ij} \bar D_k \beta^k - 2 \alpha \bar A_{ik} \bar A^k {}_j 
+ \alpha \bar A_{ij} K \nonumber \\
&& + e^{- 4 \phi} \Big[ - 2 \alpha \bar D_i \bar D_j \phi + 4 \alpha \bar D_i \phi \bar D_j \phi \nonumber \\
& & ~~~~~~~ + 4 \bar D_{(i} \alpha \bar D_{j)} \phi - \bar D_i \bar D_j \alpha 
\nonumber\\ 
& & ~~~~~~~ + \alpha (\bar R_{ij} - 8\pi S_{ij}) \Big]^{\rm TF} \\
\part_n \phi & = & \frac{1}{6} \bar D_k \beta^i - \frac{1}{6} \alpha K \\
\part_n K & = & \frac{\alpha}{3} K^2 + \alpha \bar A_{ij} \bar A^{ij} 
- e^{- 4 \phi}  ( \bar D^2 \alpha + 2 \bar D^i \alpha \bar D_i \phi ) \nonumber \\
& & + 4 \pi \alpha (\rho + S) \\
\part_n \bar \Lambda^i & = & \bar \gamma^{jk} \Dflat_j \Dflat_k \beta^i 
+ \frac{2}{3} \Delta \Gamma^i \bar D_j \beta^j + \frac{1}{3} \bar D^i \bar D_j \beta^j \nonumber \\
&  & - 2 \bar A^{jk} ( \delta^i{}_j \partial_k \alpha - 6 \alpha \delta^i{}_j \partial_k \phi
- \alpha \Delta \Gamma^i_{jk} ) \nonumber \\
& & - \frac{4}{3} \alpha \bar \gamma^{ij} \partial_j K - 16 \pi \alpha \bar \gamma^{ij} S_j \ ,
\end{eqnarray}
\end{subequations}
where $\alpha$ is the lapse function, $\Dflat_i$ denotes a covariant derivative that is built from the background connection $\Gammaflat^i_{jk}$ and the superscript ${\rm TF}$ denotes the trace-free part.  The matter sources $\rho$, $S_i$, $S_{ij}$ and $S$ denote the density, momentum density, stress, and the trace of the stress as observed by a normal observer, respectively, and are defined by
\begin{subequations}
\begin{eqnarray}
\rho & \equiv & n_a n_b T^{ab}, \\
S_i & \equiv & - \gamma_{ia} n_b T^{ab}, \\
S_{ij} & \equiv & \gamma_{ia} \gamma_{jb} T^{ab}, \\
S & \equiv & \gamma^{ij} S_{ij}.
\end{eqnarray}
\end{subequations}
Here
\begin{equation}
n_a = (-\alpha,0,0,0)
\end{equation}
is the normal one-form on a spatial slice, and $T^{ab}$ is the stress-energy tensor.

We compute the Ricci tensor $\bar R_{ij}$ associated with $\bar \gamma_{ij}$ from
\begin{eqnarray} \label{ricci}
\bar R_{ij} & = & - \frac{1}{2} \bar \gamma^{kl} \Dflat_k \Dflat_l \bar \gamma_{ij} + 
\bar \gamma_{k(i} \Dflat_{j)} \bar \Lambda^k + \Delta \Gamma^k \Delta \Gamma_{(ij)k} \nonumber \\
& & + \bar \gamma^{kl} \left( 2 \Delta \Gamma^m_{k(i} \Delta \Gamma_{j)ml} 
+ \Delta \Gamma^m_{ik} \Delta \Gamma_{mjl} \right).
\end{eqnarray}

The Hamiltonian constraint takes the form
\begin{eqnarray} \label{Ham}
{\mathcal H} & \equiv & \frac{2}{3} K^2 - \bar A_{ij} \bar A^{ij} + e^{- 4 \phi} ( \bar R - 8 \bar D^i \phi \bar D_i \phi - 8 \bar D^2 \phi) \nonumber \\
& & - 16 \pi \rho \nonumber \\
& = & 0,
\end{eqnarray}
and the momentum constraints can be written as
\begin{eqnarray}
{\mathcal M}^i & \equiv & e^{-4\phi} \Big( 
\frac{1}{\sqrt{\bar \gamma}} \Dflat_j(\sqrt{\bar \gamma} \bar A^{ij})
+ 6 \bar A^{ij} \partial_j \phi \nonumber \\ 
& & ~~~~~ - \frac{2}{3} \bar \gamma^{ij} \partial_j K + \bar A^{jk} \Delta \Gamma^i_{jk} 
\Big) - 8 \pi S^i
\nonumber \\
& = & 0.
\end{eqnarray}

\subsection{The Proca equations}
The Proca field is split into  its scalar $\Phi$, 3-vector $a_{i}$ potential and a three-dimensional ``electric" ${\bf E}$ and ``magnetic" ${\bf B}$ fields. The evolution equations for the complex Proca field take the form~\cite{Zilhao:2015tya}:
\begin{eqnarray}
\label{vectorpotential}\partial_{t}a_{i}&=&-\alpha(E_{i}+D_{i}\Phi)-\Phi D_{i}\alpha + \mathcal{L}_{\beta}a_{i},\\
\partial_{t}E^{i}&=&\alpha(KE^{i}+\mu^{2}a^{i}+
\epsilon^{ijk}D_{j}B_{k})\nonumber\\
&&-\epsilon^{ijk} B_{j}D_{k}\alpha+\mathcal{L}_{\beta}E^{i},\\
\partial_{t}B^{i}&=&\alpha(KB^{i}-\epsilon^{ijk}D_{j}E_{k})\nonumber\\
&&+\epsilon^{ijk} E_{j}D_{k}\alpha+\mathcal{L}_{\beta}B^{i},\\
\label{scalarpotential}\partial_{t}\Phi&=&-a^{i}D_{i}\alpha+\alpha(K\Phi-D_{i}a^{i})+\mathcal{L}_{\beta}\Phi.
\end{eqnarray}

The stress-energy tensor of the Proca field reads~\cite{Brito:2015pxa}
\begin{eqnarray} 
T_{ab}&=& -\mathcal{F}_{c(a}  \bar {\mathcal{F}}_{b)}^{\,\,c}-\frac{1}{4}g_{ab}\mathcal{F}_{cd}\bar{\mathcal{F}}^{cd} 
\nonumber \\
&+& \mu^2 \left[
\mathcal{A}_{(a}\bar{\mathcal{A}}_{b)}-\frac{1}{2}g_{ab}\mathcal{A}_c\bar{\mathcal{A}}^{c}
\right]\,,
\end{eqnarray}
from which we can compute the source terms of the Einstein equations. These are given by 
\begin{eqnarray}
&&8\pi \rho = \gamma_{ij}(E^{i}E^{j}+B^{i}B^{j})+
\mu^{2}(\Phi^{2}+\gamma^{kl}a_{i}a_{j})\ ,\\
&&4\pi(\rho+S) = \gamma_{ij}(E^{i}E^{j}+B^{i}B^{j})+2\mu^{2}\Phi^2\ ,\\
&&4\pi\biggl(S_{ij}-\frac{S}{3}\gamma_{ij}\biggl) = -\gamma_{ik}\gamma_{jl}(E^{k}E^{l} + B^{k}B^{l})\nonumber\\
&&\,\quad\quad +\frac{\gamma_{ij}}{3}(E^{2}+B^{2})+\mu^{2}a_{i}a_{j}-
\mu^{2}\frac{\gamma_{ij}}{3}\gamma^{kl}a_{k}a_{l}\ ,\ \ \ \\
&&4\pi j^{i} = \gamma^{li}\epsilon_{ljk}E^{j}B^{k}+
\mu^{2}\Phi\gamma^{ik}a_{k}\ .
\end{eqnarray}

We can define the ``Gauss" constraint as
\begin{equation}\label{Gauss}
G=D_{i}E^{i}+\mu^{2}\Phi=0 \ .
\end{equation}

Since for the evolutions reported in this paper we are assuming spherical symmetry, we only need to
consider the radial component of the vectors and the magnetic field ${\bf B}=0$.  

\section{Proca star solution and Initial Data}
\label{sec3}
Proca stars were obtained in~\cite{Brito:2015pxa} as stationary solutions to 
the model described by the action~\eqref{model}. Here we shall focus on spherically symmetric Proca 
stars, that we now briefly review, and that will be taken as the initial data 
for the time evolutions performed in the next section.

\subsection{The stationary solutions}
We consider a spherically symmetric line element 
\begin{eqnarray}
\label{ansatz1}
ds^2=-e^{2F_0} dt^2+e^{2F_1}\left[dr^2+r^2 (d\theta^2+  \sin^2\theta  d\varphi^2) \right] ,
\end{eqnarray} 
where $F_0,F_1$ are radial functions and $r,\theta,\varphi$ correspond  to 
isotropic coordinates.  The Proca field ansatz is given in terms of another two 
real functions $(V,H_1)$ which depend also on $r$
\begin{eqnarray}
\label{Paxial}
\mathcal{A}=e^{-iw t}\left( iVdt+
\frac{H_1}{r}dr   
\right) \ ,
  \ \
\end{eqnarray}
where $w>0$ is the frequency of the field.
The Einstein-Proca equations are solved with appropriate boundary conditions, which are compatible
with an approximate construction of the solutions
on the boundary of the domain of integration. This ansatz corresponds to the conventions in~\cite{Brito:2015pxa} for the axi-symmetric case, specialized to spherical symmetry. Illustrative examples of the four radial functions above, for two spherical Proca star solutions, can be found in Fig. 2 of Ref.~\cite{Brito:2015pxa} (but in a different radial coordinate). Some details on the construction of static Proca star solutions are given in Appendix~\ref{appendixA}.


The ADM mass $vs.$ Proca field frequency diagram of the solutions
is shown in Fig. 1, where we highlight the sample of numerical solutions 
that will be used in the evolutions reported in Section~\ref{results}. The 
corresponding frequency, ADM mass, Noether charge $Q$ ($cf.$ Eq.~\eqref{noetherq}), magnitude of the Proca ``electric" potential at the origin
($cf.$ Eq.~\eqref{propot}) and 
also the apparent horizon (AH) mass resulting from the unperturbed evolution 
(see next Section), are summarized in Table I.

\begin{figure}[t]
\centering
\includegraphics[height=2.45in]{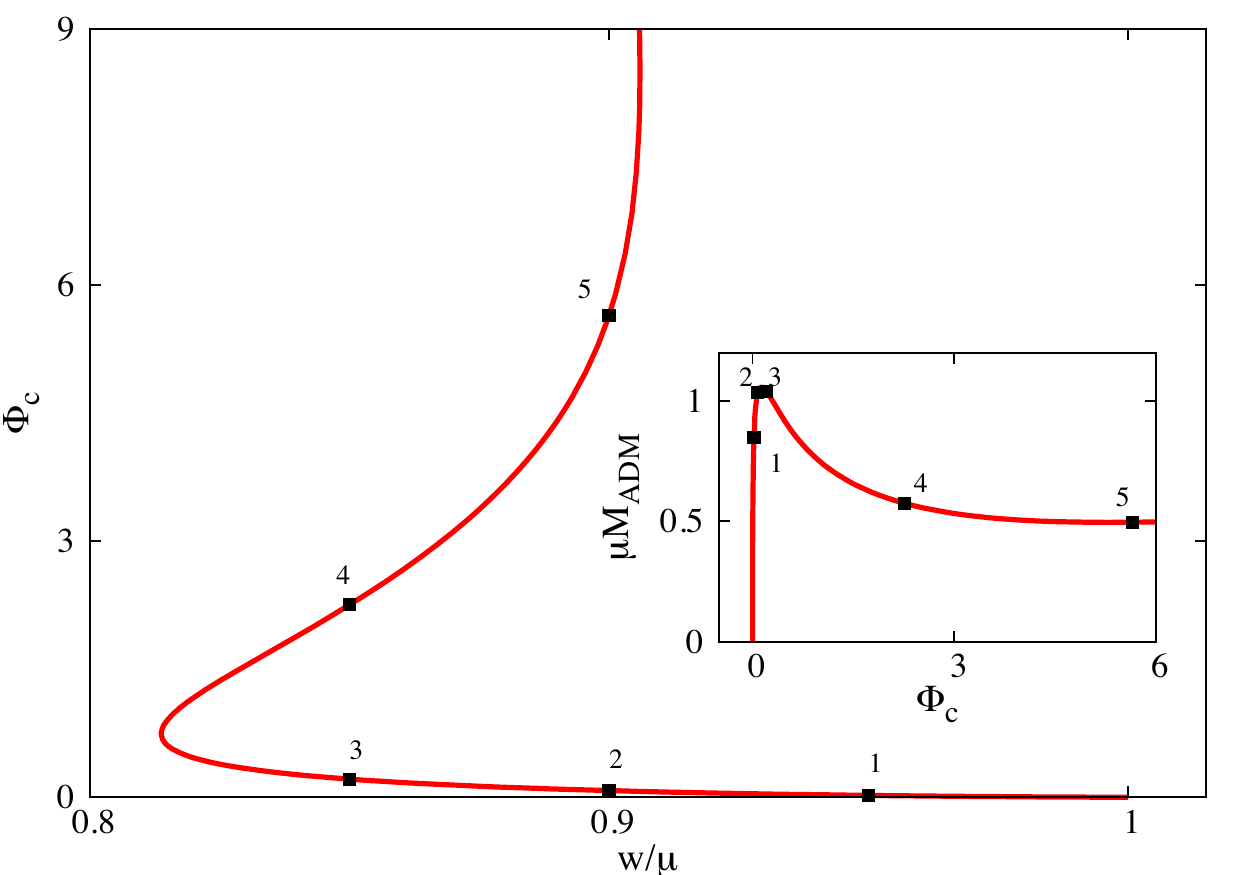}  
\caption{Magnitude of the Proca ``electric" potential at the origin, $\Phi_c(r=0)$ $vs.$ the frequency.  
The five highlighted points correspond to the same as in Fig.~\ref{fig1}. The 
inset shows the ADM mass as a function of $\Phi_c(r=0)$.
}
\label{fig1b}
\end{figure}

 \begin{table}[h!]
\caption{Spherically symmetric Proca star models.}
\label{tab:mod1}
\begin{ruledtabular}
\begin{tabular}{cccccc}
Model&$\omega/\mu$&$\mu M_{\rm ADM}$&$\mu^2Q$&$\Phi_c(r=0)$&$\mu M_{\rm{AH}}$\\
\hline
PS1&0.95&0.849&0.864 &0.0214&-\\
PS2&0.90&1.036&1.063 &0.0779&-\\
PS3&0.85&1.039&1.065   &0.2121&1.045\\
PS4&0.85&0.576&0.504&2.2547&0.578\\
PS5&0.90&0.496&0.412 &5.6473&0.489\\

\end{tabular}
\end{ruledtabular}
\end{table}

Proca stars carry a conserved Noether charge, $Q$, which is a measure of 
the number of Proca particles in the Bose-Einstein condensate. It is given by integrating the temporal component of the conserved current associated to the global $U(1)$ symmetry on a spacelike surface:
\begin{equation}
Q=\frac{i}{2} \int_\Sigma d^3 x\sqrt{-g}\left(\bar{\mathcal{F}}^{t\beta}\mathcal{A}_\beta-\mathcal{F}^{t \beta}\bar{\mathcal{A}}_\beta\right) \ .
\label{noetherq}
\end{equation}
Multiplying this 
number by the mass of one individual Proca particle, $\mu$, one obtains a measure 
of the total energy excluding binding energy. A comparison with the ADM mass of 
each solution then discriminates solutions with a positive binding energy, 
$M_{\rm ADM}<\mu Q$, from the ones with a negative binding energy, or excess energy, 
$M_{\rm ADM}>\mu Q$. The latter, naturally, only occur in the unstable branch. The 
separation point (zero binding energy) turns out to occur very close to (but not 
exactly at) the solution with the minimum frequency (see Fig. 1 
in~\cite{Brito:2015pxa}). This means that the classification of our five 
illustrative solutions is, from perturbation theory plus binding energy 
considerations, as follows:
\begin{description}
\item[i)] Models 1 and 2: {\bf stable} (which implies positive binding energy);
\item[ii)] Model 3: {\bf unstable} with {\bf positive} binding energy;
\item[iii)] Model 4 and 5: {\bf unstable} with {\bf negative} binding energy/excess energy.
\end{description}

The translation between the four radial functions above, $F_0,F_1,V,H_1$, and 
the initial value for the Proca field variables described in 
Eqs~(\ref{vectorpotential})-(\ref{scalarpotential}) is given as follows:
\begin{eqnarray}
\Phi&=&-n^{\mu}\mathcal{A}_{\mu}\  , \label{propot}\\
a_{i}&=&\gamma^{\mu}_{i}\mathcal{A}_{\mu}\ ,\\
E^{i}&=&-i\,\frac{\gamma^{ij}}{\alpha}\,\biggl(D_{i} (\alpha\Phi)+\partial_{t}a_{j}\biggl) \ .
\end{eqnarray}
Then, from (\ref{Paxial}), we obtain
\begin{eqnarray}
\Phi&=&-i\frac{V}{\alpha}\ , \\
a_{r}&=&\frac{H_{1}}{r}\ ,\\
E^{r}&=&i\frac{\gamma^{rr}}{\alpha}\,\biggl(D_{r} V+\omega\frac{H_{1}}{r}\biggl) \ .
\end{eqnarray}

In the evolutions to be discussed below, one of the observables of interest is 
the energy in the Proca field as well as the energy in the black hole, in case 
one forms. 
A natural way to assign well defined such energies in a stationary, 
asymptotically flat spacetime is by using Komar integrals. The Komar mass at 
infinity coincides with the ADM mass, $M_{\rm ADM}$. By using Gauss's law, this 
mass can be computed as a spatial volume integral of the ``Komar energy density" 
along a spacelike slice from a horizon, in case one exists, up to spatial 
infinity, plus the horizon contribution, $M_{\rm AH}$, which is, again, computed 
as a Komar integral, $M_{\rm ADM}=M_{\rm AH}+E_{\rm PF}$, 
where~\cite{Herdeiro:2016tmi}:
\begin{equation}
E_{\rm PF}=-\int_{\Sigma}drd\theta 
d\varphi\left(2T^t_t-T_\alpha^\alpha\right)\alpha\sqrt{\gamma} \ .
\label{energy}
\end{equation}   
In our evolutions we will use $E_{\rm PF}$ with the volume integral computed from the AH (origin) to spatial infinity, in case an AH is present (absent). For the horizon mass we shall use the irreducible mass computed from the horizon area, since the black hole that forms has no electric charge or rotation. 

%

\section{Numerics} 
\label{sec:numerics}

As mentioned before, the numerical relativity code we use to perform the evolution of the Einstein-Proca system is 
based on the code originally presented in \cite{baumgarte2013numerical}. The most salient feature of this code, as
compared to other numerical relativity codes, is the fact that the equations are implemented and solved using spherical 
polar coordinates. The time update of the different systems of evolution equations (Einstein and Proca) is therefore 
done using the same type of techniques we have extensively used in our previous work. The interested reader is addressed 
in particular to Refs.~\cite{Montero:2012yr,Sanchis-Gual:2015bh,Sanchis-Gual:2015sms} for complete details. 

As a summary, we indicate here that the evolution equations are integrated using 
the second-order Partially Implicit Runge-Kutta (PIRK) method developed by \cite{Isabel:2012arx,Casas:2014}.  This 
method allows to handle the singular terms that appear in the evolution equations due to our choice 
of curvilinear coordinates. For the simulations presented in this paper, the original code had to be upgraded to
account for the Proca field. Therefore, the evolution scheme for this field is the only feature which is new with respect
to previous versions of the code. Therefore, it is worth discussing in some more detail our specific implementation.

Following \cite{Isabel:2012arx,Casas:2014} we cast the Einstein-Proca system of PDEs as   
\begin{System}
u_t = \mathcal{L}_1 (u, v) \ , \\
v_t = \mathcal{L}_2 (u) + \mathcal{L}_3 (u, v) \ ,
\label{e:system}
\end{System}
where $\mathcal{L}_1$, $\mathcal{L}_2$ and $\mathcal{L}_3$ represent general non-linear 
differential operators. In the second-order PIRK method the time update from time $t^n$ to $t^{n+1}$ is done according to
the following two-step algorithm:
\begin{System}
	u^{(1)} = u^n + \Delta t \, L_1 (u^n, v^n)º \ ,  \\
	v^{(1)} = v^n + \Delta t \left[\frac{1}{2} L_2(u^n) +
          \frac{1}{2} L_2(u^{(1)}) + L_3(u^n, v^n) \right] \ ,
\end{System}
\begin{System}
	u^{n+1}  = \frac{1}{2} \left[ u^n + u^{(1)} 
+ \Delta t \, L_1 (u^{(1)}, v^{(1)}) \right] \ , \\
	v^{n+1}  =   v^n + \frac{\Delta t}{2} \left[ 
L_2(u^n) + L_2(u^{n+1}) \right.  \\ 
\left. \hspace{2.5cm} +  L_3(u^n, v^n) + L_3 (u^{(1)}, v^{(1)}) \right] \ ,
\end{System}
where $L_1$, $L_2$ and $L_3$ are the corresponding discrete operators. 
The explicit form of these operators for the (BSSN) Einstein evolution
equations are shown in \cite{baumgarte2013numerical}. Regarding the Proca field,
each time step starts by evolving explicitely the vector potential $a_{i}$, $i.e.$, all the
source terms of the evolution equations of these variables are
included in the $L_1$ operator.

We then evolve the scalar potential $\Phi$. More precisely, the corresponding $L_2$ and $L_3$
operators associated with the evolution equations for $a_{i}$ are
\begin{eqnarray}
        L_{2( \Phi)} &=& -a^{i}D_{i}\alpha - \alpha D_{i}a^{i} \label{L2A}, \\
        L_{3( \Phi)} &=& \alpha\,K\Phi + \mathcal{L}_{\beta}\Phi. 
\end{eqnarray} 
The components of the electric field $E^{i}$ are evolved partially implicitly, using the updated values of $\Phi$ and $a_{i}$. 
The corresponding operators are
\begin{align}
        L_{2(E^i)} &= \alpha(KE^{i}
        +\mu^{2}a^{i}+\epsilon^{ijk}D_{j}B_{k})\nonumber\\
&-\epsilon^{ijk} B_{j}D_{k}\alpha,\\   
	\label{L3Lambda}
        L_{3(E^i)} &= \mathcal{L}_{\beta}E^{i} .
\end{align}

Finally, we note that the spatial derivatives in the spacetime evolution are computed using a fourth-order, centered, 
finite-difference approximation on a logarithmic grid (see~\cite{Sanchis-Gual:2015sms}). Moreover, as customary in 
grid-based numerical relativity codes, we use fourth-order Kreiss-Oliger dissipation to avoid high frequency noise 
appearing near the outer boundary. Details about the convergence properties and accuracy of the code are given in
Appendix~\ref{appendix}.

In the numerical simulations we have used a logarithmic radial grid that extends from the origin to $r = 50$ and uses a minimum resolution close to the origin of
$\Delta r = 0.0125$.
\section{Results}
\label{results}

\subsection{Unperturbed evolutions}

We start by evolving the five spherically symmetric Proca star models described in 
Table~\ref{tab:mod1} without any additional perturbation (apart from the 
unavoidable discretization error).

\begin{figure}
\centering
\includegraphics[height=2.45in]{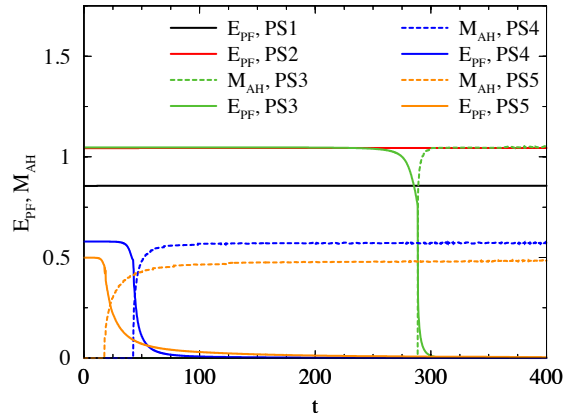} 
\caption{Time evolution of the Proca field energy and the apparent horizon mass for the unperturbed models 1- 5. In this and remaining time series, the time coordinate is given in units of $\mu$.}
\label{fig2}
\end{figure}

\begin{figure}
\centering
\includegraphics[height=2.45in]{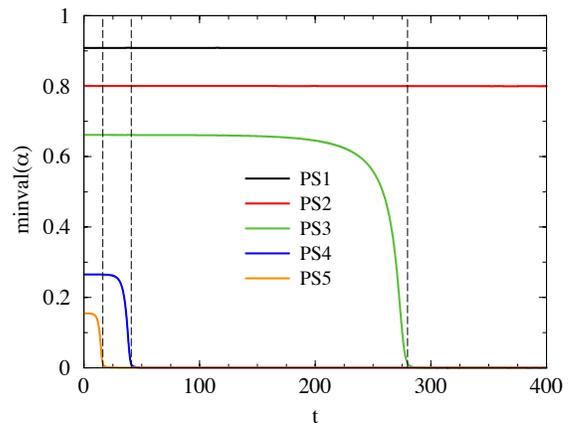}
\caption{Time evolution of the minimal value of the lapse for all unperturbed models. The vertical dashed lines indicate
the time of formation of an AH for models 3, 4, and 5.
}
\label{fig3}
\end{figure}

 In Fig.~\ref{fig2} we plot the time evolution of the Proca field energy 
computed from equation~(\ref{energy}) and the irreducible mass of the AH, when 
the latter is present.  The figure shows that the energy 
remains constant for models 1 and 2, hence showing their stability, at least in 
the absence of larger perturbations. For the other three models, the discretization 
error is sufficient to trigger the collapse of the solutions. At some point in 
the evolution an AH forms. This is confirmed in Fig.~\ref{fig3}, 
where the time evolution of the minimum of the lapse function is plotted, showing 
it tends to zero, a signature of AH formation. The precise instants those horizons are found in our code
are indicated by the three vertical dashed lines in the figure. Subsequently,  all Proca field 
energy evolves towards being absorbed by the black hole. The results in Figs.~\ref{fig2}-\ref{fig3} also 
indicate that decay is progressively faster as we move away from the threshold 
of stability solution (maximal ADM mass) in Fig.~\ref{fig1}. These results match 
precisely the expectation from perturbation theory in the frequency 
domain~\cite{Brito:2015pxa}. 

\begin{figure}[h!]
\centering
\includegraphics[height=2.45in]{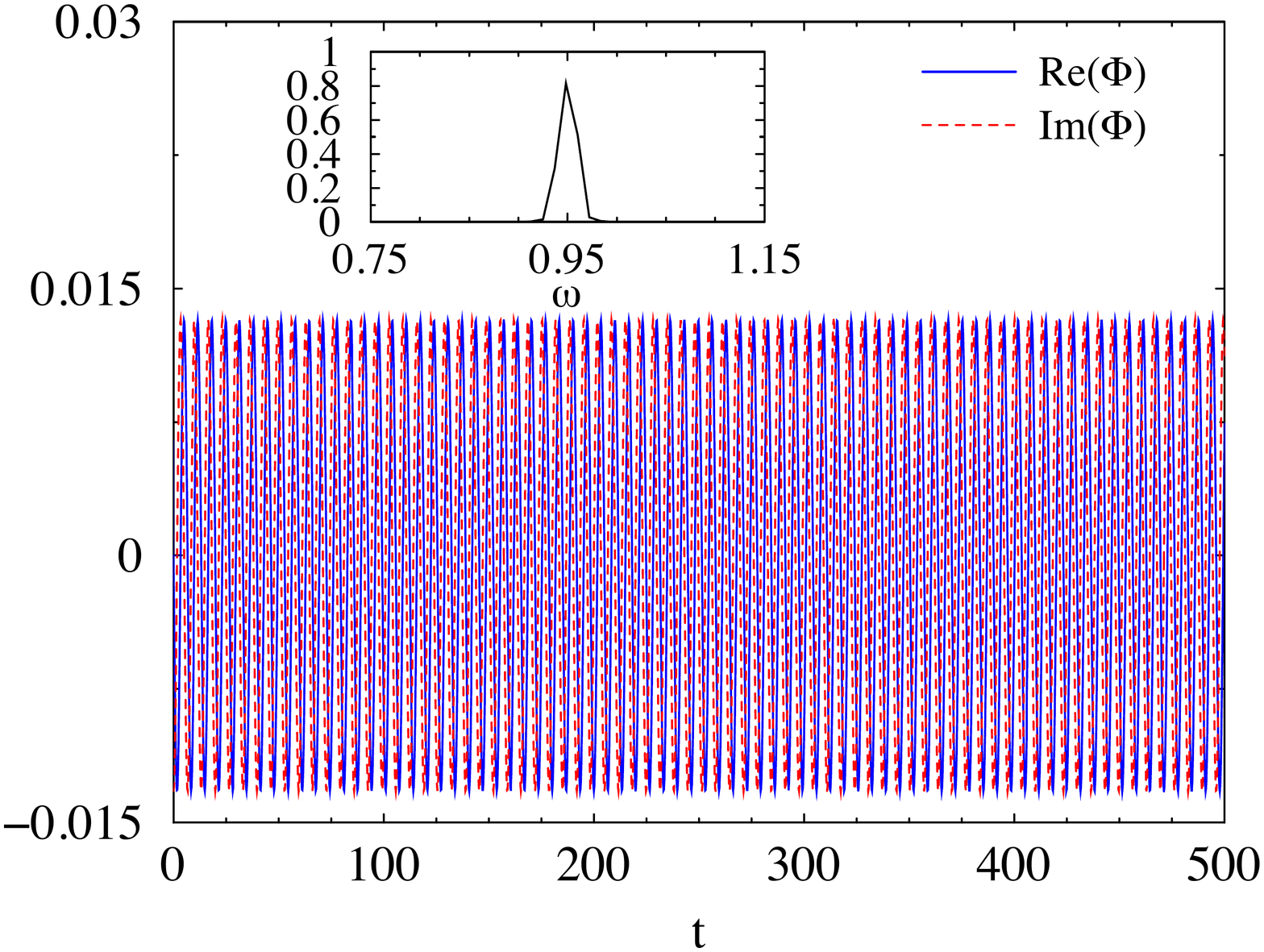}\\\includegraphics[height=2.45in]{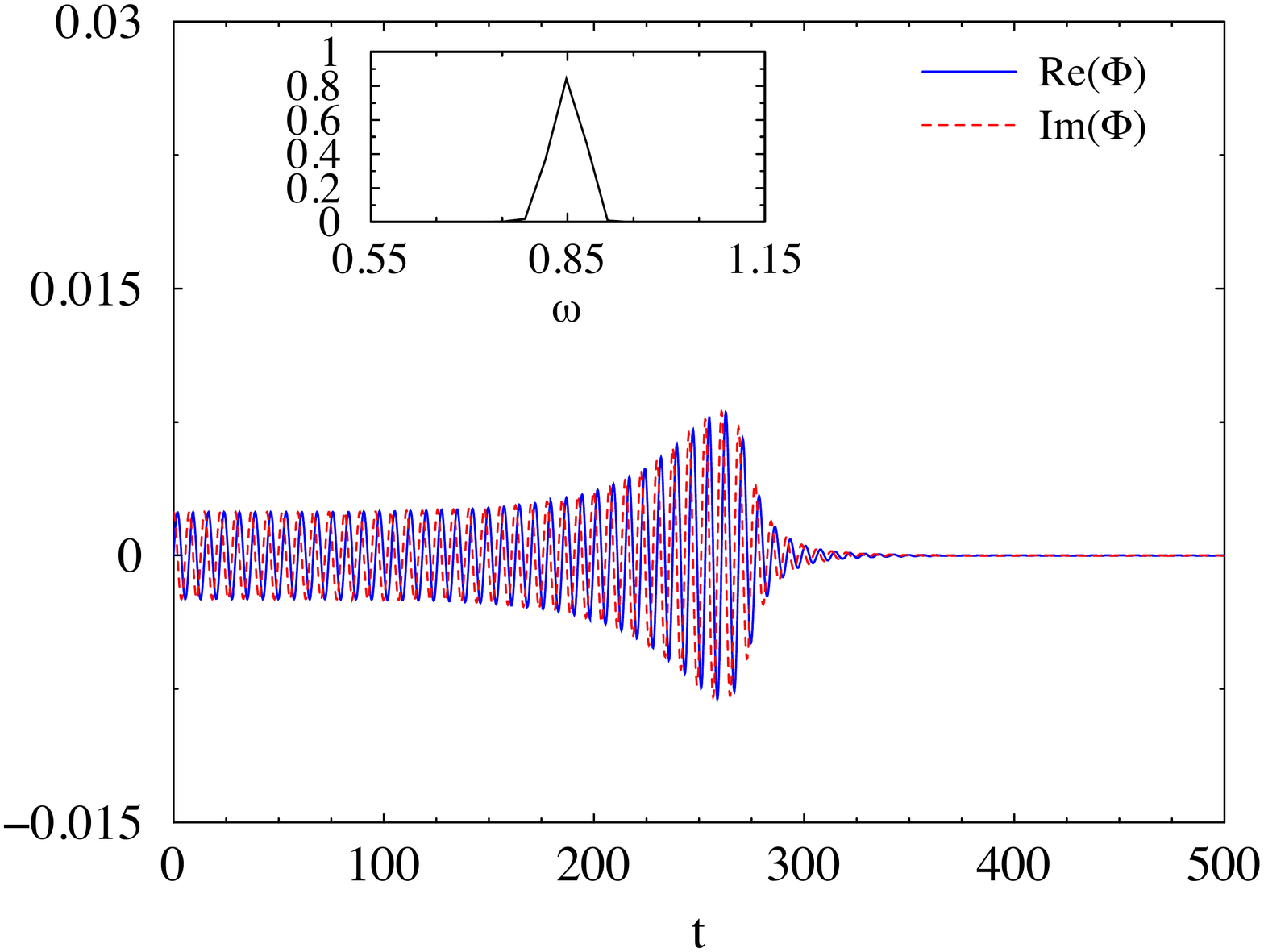}\\
\includegraphics[height=2.45in]{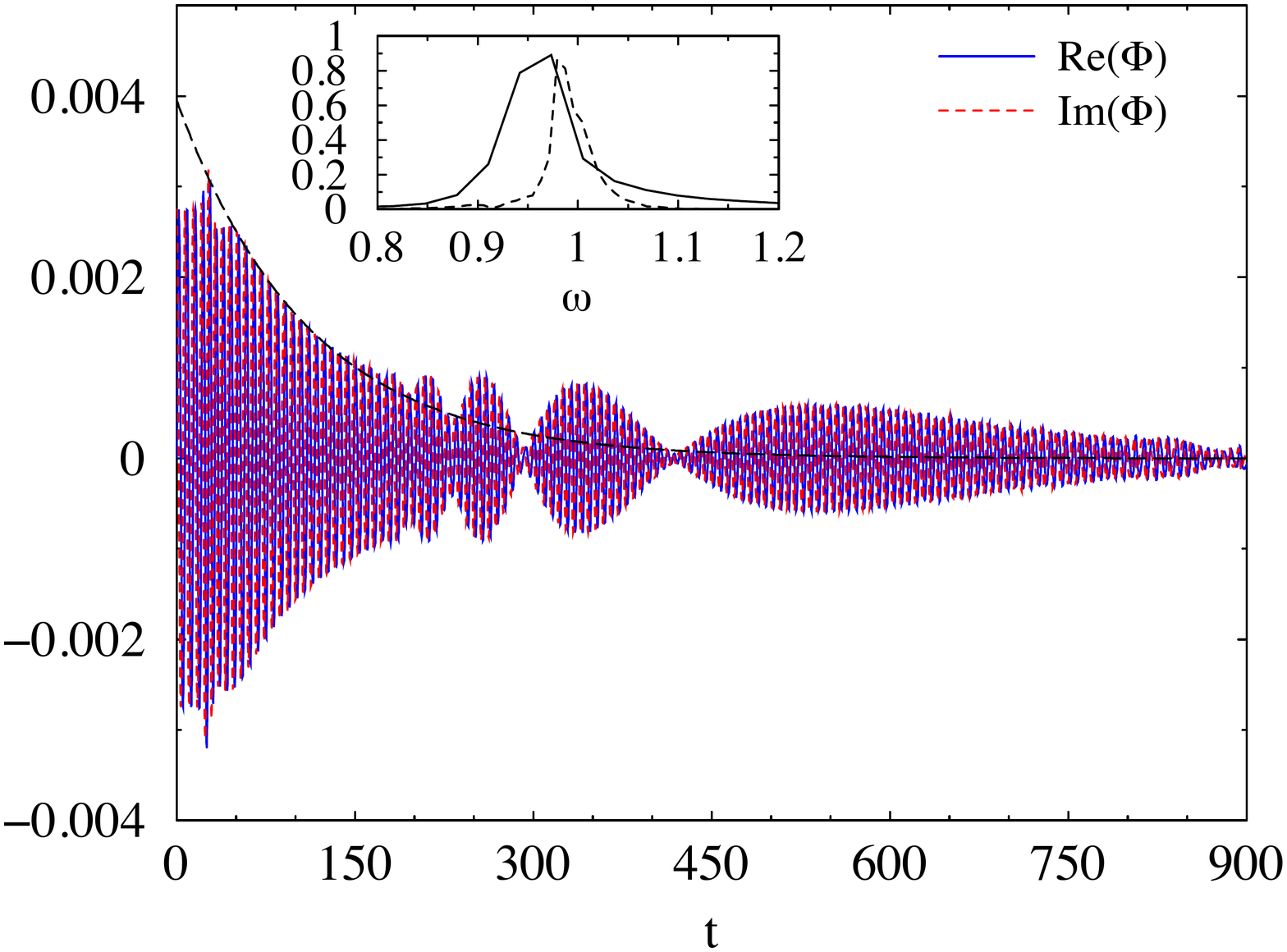}
\caption{Time evolutions of the amplitude of the scalar potential extracted at $r=5$ for models 1 (top manel), 3 (middle panel), and
5 (bottom panel). The oscillation frequencies shown in the insets of the top and middle panel are in good agreement with the corresponding $\omega$, in units of $\mu$, of 
models 1-3 (see Table~\ref{tab:mod1}). In the bottom panel, the initial decay of the wig is fitted by the enveloping function $e^{-\omega_{I}\,t}$, where for model 5, $\omega_{I}/\mu\sim \,0.009$.}
\label{fig4}
\end{figure}

Let us now turn our attention to Fig.~\ref{fig4}. This figure shows the time evolution of the amplitude 
of the scalar potential at some extraction radius and for three representative models. The top panel (model 1) shows constant amplitude oscillations, with a frequency matching that of the Proca stars, as expected for a solution that is not changing in time. The Fourier transform of the time series reveals (inset) that the corresponding frequency is that of the Proca star. The middle panel (model 3) shows oscillations that remain constant up to $t\sim 150$, then grow and then decay to zero. At about $t\sim 300$ the AH is formed (Fig.~\ref{fig3}) and shortly afterwords no noticeable Proca field remains outside the horizon. Again, the inset shows that the frequency in the time series is the one of the original Proca star.
The bottom panel (model 5 and likewise for model 4) shows  a different behaviour from model 3 (middle panel), 
in spite of both leading to black hole formation. When the black hole is 
formed, a more significant part 
of the Proca field lingers for a longer period of time. This is already 
visible comparing the Proca field energy curves for models 4 or 5 and model 3, in 
Fig.~\ref{fig2}. Moreover, the \textit{beating pattern} observed in 
the bottom panel of Fig.~\ref{fig4} shows that more than one significant frequency is present. This is indeed confirmed by a Fourier analysis (inset). 
Such an analysis can be compared with the frequencies of the first few monopole quasi-bound state of a Proca field on a Schwarzschild BH~\cite{Rosa:2011my} -- Table~\ref{tab:mod3}\footnote{We display the frequencies for both the fundamental mode ($n=0$) and first overtones ($n=1,2,3$), where $n$ is the excitation quantum number~\cite{Rosa:2011my}  and for both $\mu M_{\rm ADM}=0.5$ and $\mu M_{\rm ADM}=0.45$. Since  the original Proca star has mass $\mu M_{\rm ADM}=0.496$, the mass of the Schwarzschild black hole resulting from the collapse of model 5, around which a small Proca field remnant  lingers, is estimated to be in between these two values. We would like to thank J. G. Rosa for providing us with these frequencies.} -- and yields the following conclusions:
 \begin{table}[h!]
\caption{Frequencies for spherical (monopole) quasibound states of a Proca field (mass $\mu$) around a Schwarzschild black hole (mass $M_{\rm ADM}$).}
\label{tab:mod3}
\begin{ruledtabular}
\begin{tabular}{ccc}
Mode & $w/\mu$ ($\mu M_{\rm ADM}=0.45$) & $w/\mu$ ($\mu M_{\rm ADM}=0.5$) \\
\hline
  $n=0$ & $0.970-0.018i$   & $0.966-0.025i$\\
$n=1$&  $0.986-0.005i$& $0.984-0.008i$ \\
$n=2$&$0.992-0.002i$ & $0.991-0.003i$ \\
$n=3$&$0.995-0.001i$ & $0.994-0.002i$ \\
\end{tabular}
\end{ruledtabular}
\end{table}
\begin{description}
\item[$\bullet$] The initial part of the time series ($t=0-200$, black solid line in the inset) contains the initial Proca star frequency $w/\mu=0.9$; its frequency range is centered around $w/\mu=0.96-0.97$,  which matches the fundamental mode frequency, whose real part is $w\sim 0.966-0.970$, for $\mu M_{\rm ADM}=0.5$ or $0.45$ respectively (Table~\ref{tab:mod3}). The peak frequency is slightly higher than the central one, suggesting that an overtone (rather than the fundamental mode) dominates the signal. This is consistent with the enveloping (dashed black) fit curve presented which has $\omega_{I}/\mu\sim \,0.009$, a value closer to the imaginary part of the first overtone than that of the fundamental mode.
\item[$\bullet$]  The later part of the time series ($t=200-900$, dashed line in the inset) is narrower and centered around $w/\mu\sim 0.98-0.99$. The absence of the fundamental mode is expected as its lifetime is $\tau\sim 1/|\omega_I| \lesssim 100$. The signal is then composed essentially by the overtones which are longer lived than the fundamental mode, and have frequencies in the range  $w/\mu\sim 0.98-1$ -- Table~\ref{tab:mod3}. The beating pattern present shows the existence of more than one overtone.
\end{description}

These lingering Proca modes resemble the 
scalar quasi-bound states discussed 
in~\cite{Barranco:2012qs,Sanchis-Gual:2015bh} and the beating pattern resembles 
that observed in superpositions of quasi-bound states of a scalar field around 
Schwarzschild black holes  - see $e.g.$ \cite{Okawa:2014nda,Barranco:2013rua}. 

We recall that test field, frequency domain studies have shown that the Proca field exhibits 
quasi-bound states, $i.e.$ solutions which are 
localized within the vicinity of a black hole~\cite{Galtsov:1984ixy}. 
These quasi-bound states, as in the scalar case,
have complex frequencies and the imaginary part corresponds to the 
decay rate. In the low mass limit, the fundamental modes of these states, 
are consistent with a hydrogenic spectrum.
A thorough study of the quasi-bound states and the quasi-normal modes 
of the Proca field in a Schwarzschild background can be found in 
\cite{Rosa:2011my}. The nonlinear dynamics of self-gravitating Proca 
quasi-bound states in both Schwarzschild and Kerr black holes has also been 
studied numerically in \cite{Zilhao:2015tya}.
In particular, the study of Proca fields on the background of rotating black 
holes has been used to provide upper limits on the mass of the photon~\cite{Pani:2012vp}.

\subsection{Perturbed evolutions}

We turn now to the study of the dynamical behaviour of Proca stars under perturbations. To illustrate this 
situation we add a perturbation (at the few \% level) to the Proca field variables of the form $\tilde A  \rightarrow  
\tilde Ax$ where $\tilde A$ stands for $\Phi, a^r, E^r$  and $x$ is a numerical 
factor. As a remark, we observe that this type of perturbation does not change the sign of the binding energy of the Proca star. Indeed, inspection of~\eqref{noetherq} and~\eqref{energy}, shows that both these quantities scale as the square of the scaling factor. Thus, this scaling does not change the sign of their difference. 

We first take $x=1.02$. For this choice, the unstable Proca stars tend to collapse faster than the unperturbed solution with the same resolution. This is the expected result, since this perturbation increases, in particular, the central electric potential of the Proca star ($cf.$ discussion on neutron stars in the conclusions). 

\begin{figure}
\centering
\includegraphics[height=2.45in]{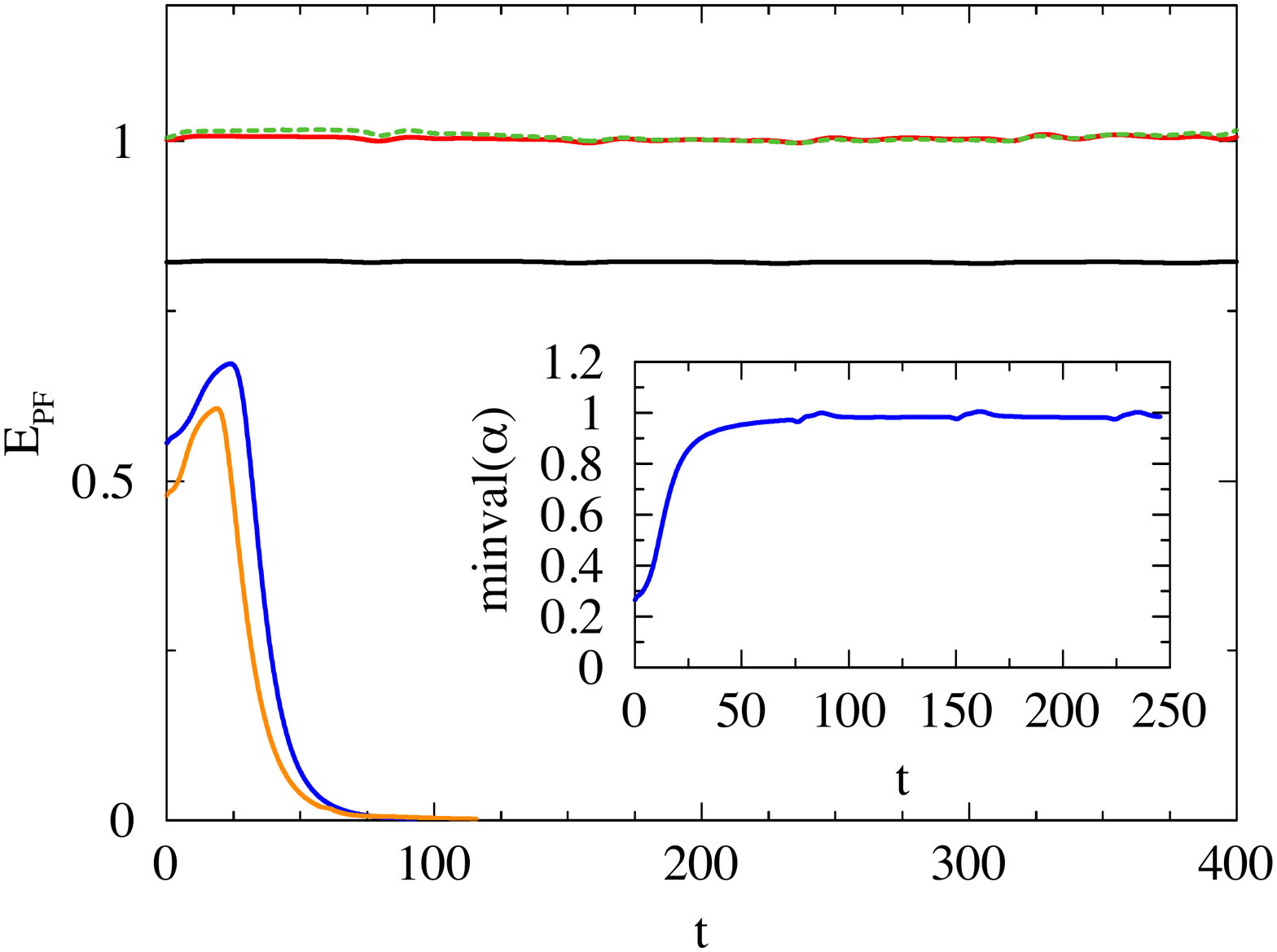}\\ 
\includegraphics[height=2.45in]{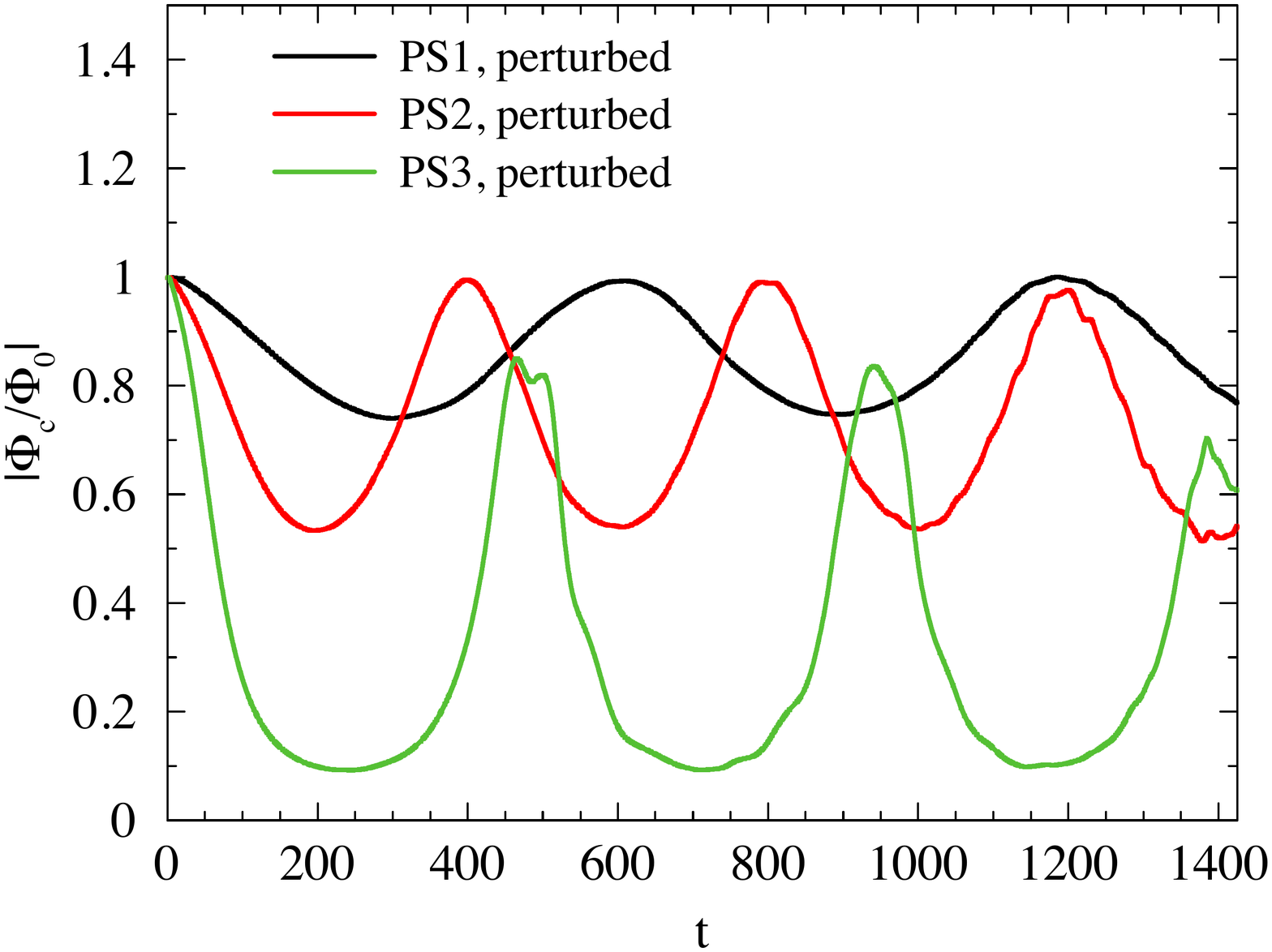}
\caption{(Top panel) Time evolution of the perturbed models. The inset shows the minimum value of the lapse of perturbed model 4.  (Bottom panel) 
Time evolution of the magnitude of the Proca ``electric" potential for models 1-3.}
\label{fig7}
\end{figure}

Next, we choose $x=0.98$. In this case, we are removing part of the Proca field and the outcome of the evolution is different for the unstable models, as exhibited in Fig.~\ref{fig7}. 
In the top panel of Fig.~\ref{fig7}, we plot the time evolution of the Proca field energy for the perturbed models (the color coding is the same as in Fig.~\ref{fig2}). Models 1 (black solid line) appears stable. Model 2 (red solid line) and 3 (green dashed line), despite fairly stable, exhibit some noticeable oscillations. We shall come back to these shortly. Models 4-5, on the other hand, show a total dispersive behaviour: the energy of the Proca field goes to zero \textit{without} AH formation; moreover, the value of the lapse function tends to unity at the origin (inset), which is the behaviour of flat spacetime. We remark that the slight increase in the Proca field energy clearly visible in the beginning of the evolution of models 4-5 is related to the evolution of the lapse function, as it adapts to the perturbation. 

A more detailed investigation of what occurs for models 1-3 can be made by 
analysing the behaviour of the electric potential at the origin, which is 
displayed in the bottom panel of Fig.~\ref{fig7}. All three models oscillate, 
but whereas the oscillations of models 1 and 2 are not damping the value of 
$\Phi_c$, the ones of model 3 are clearly decreasing its value. We recall, from 
Fig.~\ref{fig1b}, that this scalar potential at the origin defines uniquely 
static Proca solutions. The decrease of its value for model 3 can be interpreted 
as a \textit{migration} of the unstable star towards the stable branch. 
This conclusion is supported by computing the oscillation frequency of the 
perturbed models. For the perturbed model 3 it is  $\sim 0.92$, whereas for the 
perturbed model 2, it is  $\sim 0.91$. Thus, the perturbed model 3 is 
oscillating with a frequency which corresponds to a stable Proca star, very 
close to the one of model 2, which is in the stable branch. We recall that models 4 and 5 have negative binding 
energy (excess energy), unlike solution 3. It is therefore expectable they could 
have a different evolution and disperse away. This is precisely what we observe 
in these perturbed evolutions. 

To summarize, the perturbed evolutions indicate that, depending on the perturbation, but also on the sign of the binding energy, Proca stars can be subject to three different outcomes: collapse, dispersion/fission or migration.

\section{Discussion}
\label{secconclusions}

In this paper we have used numerical relativity techniques to study the 
stability and evolution of Proca stars. These are macroscopic, self-gravitating 
Bose-Einstein condensates composed by a massive, complex, vector 
field~\cite{Brito:2015pxa}. Considering the energy of these solutions, together with 
perturbative stability studies, leads to three different classes of solutions: $(i)$ Stable solutions (our models 1 and 2); $(ii)$ Unstable solutions with positive binding energy (our model 3); $(iii)$ Unstable solutions with negative binding energy or excess energy (our models 4 and 5). Here, we have shown that the fate of the unstable solutions depends not only on the type of solutions but \textit{also} on the type of perturbation. Our results are summarized in Table \ref{tab:summary} and as follows.

When we evolve in time the configurations with no perturbations we observe the following behaviours: 
Models 1 and 2 evolve in time with almost no changes. The Proca field 
oscillates and the spacetime remains stationary.
Configurations 3 collapses to form a black hole without a noticeable Proca field remnant, thus yielding a vacuum spacetime. For unstable 
configurations 4-5, however, even though a black hole is formed after the collapse a noticeable set of Proca 
quasi-bound states remain in its vicinity.
  
When we apply a perturbation, multiplying the fields by a constant of order 
unity,  the stable configurations still display no evolution. Configuration 3 migrates 
towards stable states with lower mass, whereas configurations 4-5
are no more localized states and the Proca field disperses away.

\begin{widetext}

\begin{table*}[h!]
\caption{Final state of the five representative PS models considered in this paper.} 
\label{tab:summary}
\begin{ruledtabular}
\begin{tabular}{ccc|c|c}
Model&Perturbative Analysis&Binding energy&Evolution (no perturbation)&Evolution ($\times 0.98$ perturbation)\\
\hline
PS1&stable&$>0$&stable& stable\\
PS2&stable&$>0$&stable& stable\\
PS3&unstable&$>0$&collapse to BH & migration\\
PS4&unstable&$<0$&collapse to BH (+ noticeable wig)& dispersion\\
PS5&unstable&$<0$&collapse to  BH (+ noticeable wig)& dispersion\\
\end{tabular}
\end{ruledtabular}
\end{table*}

\end{widetext}

The fates we have just described for Proca stars parallel quite closely those observed for scalar boson stars and discussed in detail in the Introduction. But it is interesting to compare this dynamics with that of the well known compact fermionic stars, namely neutron stars. 

\bigskip

In an astrophysical context, the dynamical fate of neutron stars out of equilibrium has received enormous attention, as it plays a crucial role in significant scenarios of relativistic astrophysics such as supernovae, gamma-ray bursts, or binary neutron star mergers. Moreover, as a fundamental physics problem, critical phenomena in neutron stars at the threshold of black hole formation has also been examined for perfect fluid matter (see~\cite{Noble:2016} and references therein). By triggering the collapse of initially stable models through suitable perturbations, it has been possible to
elucidate the boundary between different types of critical behavior in neutron stars and its relationship with the boundary between dispersed and bound end states~\cite{Noble:2016}. 

Radially unstable equilibrium models of neutron stars at very high densities can be built for any equation of state. Such models satisfy the condition $\partial M_{\rm ADM}/\partial \rho_{\rm c} < 0$, where $\rho_{\rm c}$
is the central rest-mass density.  Therefore, any perturbation will cause them to either collapse to a black hole or to expand to the stable branch of equilibrium models, at smaller central densities. The sign of the perturbation determines the path the unstable star will follow. These observations parallel closely what we have described for bosonic stars (both scalar and vector). 

While the two outcomes are mathematically viable, the migrating (i.e.~expanding) path is ruled out from an astrophysical viewpoint. 
In a realistic scenario, stable neutron stars, as those formed following a supernova core collapse or
an accretion-induced collapse of a white dwarf or those in X-ray binaries, can accrete matter and will secularly move towards larger central densities along the stable branch of equilibrium configurations. Beyond the maximum-mass limit collapse to a black hole will ensue. There is, however, no secular mechanism that could evolve an unstable star to the stable branch. Similarly, if Proca stars form in Nature, one may wonder if the migration dynamics we have observed is realized by any naturally occurring mechanism.

The nonlinear dynamics of a migrating relativistic spherical star was first 
investigated in~\cite{Font:2002}, using a polytropic model with a central 
rest-mass density larger than that of the maximum-mass stable model. Under the 
radial perturbation induced by the truncation error of the numerical code, the 
star was found to expand and evolve towards a stable equilibrium configuration 
with a smaller central rest-mass density but with approximately the same 
rest-mass of the perturbed star. The stable configuration is reached 
asymptotically, as the star oscillates around a central density close to that of 
a stable star with the same rest-mass. 
A similar behaviour  has been observed in numerical simulations of relativistic 
boson stars~\cite{Seidel:1990jh,Hawley:2000dt}. Contrary to the dispersive 
boson star case, for a neutron star with an ideal-fluid equation of state, the 
oscillations are gradually damped due to the dissipation of kinetic energy via 
shock heating. 


\bigskip

We would like to close with two remarks. Even though our results suggest that the dynamics and final states of PS is quite similar to SBS, the study of vector fields may yield extra value as compared to the scalar case. As a first example, 
%
consider the dynamics of Proca fields around black holes, which  has already been considered in the literature, albeit less studied than the corresponding scalar dynamics. The existence of long lived configurations around non rotating black holes has been demonstrated in the test field limit~\cite{Galtsov:1984ixy,Rosa:2011my} and in the 
full nonlinear regime solving the Einstein-Proca equations~\cite{Zilhao:2015tya}. The inherent 
difficulty of adding rotation to the black hole is also present and a full 
evolution of a rotating black hole with a Proca field is not yet available. Still, since the superradiant  instability of a Proca field has a shorter time scale than its scalar counterpart~\cite{Pani:2012bp}, this offers a promising model for analysing the non-linear development of the superradiant instability around Kerr black holes. Secondly, it has been recently argued~\cite{Conlon:2017hhi} that photons can behave near stellar mass black holes as Proca particles, in view of the effective mass induced by the galactic plasma. Thus, the study of \textit{real} Proca field interacting with black holes may describe astrophysical processes without invoking new exotic particles. 

\bigskip

\section*{Acknowledgements}

This work has been supported by the Spanish MINECO (grants AYA2013-40979-P and AYA2015-66899-C2-1-P), 
by the Generalitat Valenciana (PROMETEOII-2014-069, ACIF/2015/216), by the 
CONACyT-M\'exico, by the FCT (Portugal) IF programme, by the CIDMA (FCT) 
strategic project UID/MAT/04106/2013 and by  the  European  Union's  Horizon  2020  research  and  innovation  programme  under  the  Marie
Sklodowska-Curie grant agreement No 690904 and by the CIDMA project UID/MAT/04106/2013. Computations have been 
performed at the Servei d'Inform\`atica de la Universitat de Val\`encia and at the Blafis cluster at the University of Aveiro.

\begin{appendix}

\section{Spherically symmetric proca star solutions}
\label{appendixA}
Spherically symmetric Proca stars solutions were discussed 
in Ref. \cite{Brito:2015pxa}.
In a Schwarschild-like coordinate system the element of line is:
\begin{equation} 
ds^2=-N(\bar r) \sigma^2(\bar r) dt^2 +\frac{d\bar r^2}{N(\bar r) }+\bar r^2 (d\theta^2+\sin^2 d\varphi^2) \ .
\end{equation}
Comparison of the above line element  with the isotropic metric~\eqref{ansatz1} yields for the  
transformation
\begin{eqnarray} 
\label{transf1}
\frac{dr}{r}=\frac{d\bar r}{\bar r \sqrt{N(\bar r)}},~~~
{\rm while}~~~e^{F_0}=\sigma \sqrt{N},~~e^{F_1}=\frac{\bar r}{r}.
\end{eqnarray}
%
For the isotropic coordinates used in this paper,
the Einstein equations reduce to two second order equations for the metric functions
$F_0,F_1$ (reinserting $G_0$)
\begin{eqnarray} 
\label{eqF0}
&&
F_0''+F_0'\left(\frac{2}{r}+F_0'+F_1'\right) \\
&&
-4\pi G_0
e^{-2F_0} 
 \left[
 \left(V'+\frac{wH_1}{r}\right)^2+2e^{2F_1}\mu^2 V^2
\right]=0\ , \nonumber
\\
\label{eqF1}
&&
F_1''-F_0'F_1'+\frac{1}{r}(F_1'-F_0') \\
&&
+4\pi G_0 \mu^2 
\left(
\frac{H_1^2}{r^2}+e^{2(F_1-F_0)}V^2
\right)=0 \ , \nonumber
\end{eqnarray}
together with the constraint equation
\begin{eqnarray}
\label{constr-E}
&& 0=  F_1'^2+2F_0'F_1'+\frac{2}{r}(F_0'+F_1') +4\pi G_0 \times  \\
&& \times
\left[
e^{-2F_0}\left(V'+\frac{wH_1}{r}\right)^2
-\mu^2\left(\frac{H_1^2}{r^2}+e^{2(F_1-F_0)}V^2\right)
\right] \ .\nonumber
\end{eqnarray}
The Proca equations are:
\begin{eqnarray}
\label{eqH}
&& wV'+\frac{H_1}{r}(w^2-e^{2F_0}\mu^2)=0 \ , \\
&& V''+\frac{2V'}{r}+\left(V'+\frac{wH_1}{r}\right)(F_1'-F_0')+\frac{w}{r^2}(H_1+rH_1') \nonumber\\ 
&& -e^{2F_1}\mu^2V=0\ ,
\end{eqnarray} 
the vector potential being subject to the gauge condition
$\nabla_\alpha \mathcal{A}^\alpha=0$,
which is actually a requirement from the field equations
for the Proca field
\begin{eqnarray}
\label{gauge-cond}
 H_1'+\left(F_0'+F_1'+\frac{1}{r}\right)H_1-e^{2(F_1-F_0)}rwV=0\ .
\end{eqnarray}
For this Ansatz, the Noether charge reads
\begin{eqnarray}
\label{Q}
Q=4 \pi \int_0^{\infty} dr~e^{F_1-F_0} H_1(wH_1+r V') \ ,
\end{eqnarray} 
and the energy density
measured by a static observer is
\begin{eqnarray}
\label{Ttt}
&& -T_t^t=  \frac{1}{2}e^{-2(F_1+F_0)} \left(V'+\frac{w H_1}{r}\right)^2 \nonumber \\
&& +\frac{1}{2}\mu^2 
 \left(
 \frac{e^{2F_1}H_1^2}{r^2}+e^{-2F_0}V^2
\right) \ ,
\end{eqnarray} 
while the Komar ``energy density"  entering the integral~\eqref{energy}, is 
\begin{equation}
\label{Ttot}
2T_t^t-T_\alpha^\alpha= 
-e^{2(F_1+F_0)}
 \left(V'+\frac{w H_1}{r}\right)^2-2e^{-2F_0}\mu^2V^2 \ .
\end{equation} 
The PSs possess the following expansion at the origin $r\to 0$
\begin{eqnarray}
\label{r=0}
&&
F_0(r)=f_0+\frac{4\pi G_0}{3}e^{2(f_1-f_0)}\mu^2 v_0^2 r^2+\mathcal{O}(r^4) \ , 
\\
&&
F_1(r)=f_1-\frac{4\pi G_0}{12}e^{2(f_1-f_0)}\mu^2 v_0^2 r^2+\mathcal{O}(r^4)\ ,
\\
&&
H_1(r)=\frac{1}{3}e^{2(f_1-f_0)}v_0w r^2+\mathcal{O}(r^4) \ , \\
&&
V(r)=v_0+\frac{1}{6}e^{2(f_1-f_0)}v_0(e^{2f_0}\mu^2-w^2)r^2+ \mathcal{O}(r^4)\ , \nonumber \\
\end{eqnarray} 
where $f_1,$ $f_0$ and $v_0$ are constants fixed by the numerics.
The first terms in the expression of the solution as $r\to \infty$
read
\begin{eqnarray}
\label{inf} 
&& F_0(r)=-\frac{M_{\rm ADM}}{r}+\dots \ , \\
&&
F_1(r)=\frac{M_{\rm ADM}}{r}+\dots\  , \\
&&
H_1(r)=-c_0\frac{w}{\sqrt{\mu^2-w^2}}e^{-r\sqrt{\mu^2-w^2}}+\dots \ , \\
&& 
V(r)=c_0\frac{e^{-r\sqrt{\mu^2-w^2}}}{r}+\dots,
\end{eqnarray} 
where $c_0$ is a constant. Observe that the PSs should satisfy the bound state condition $w < \mu$. 

The solutions that smoothly interpolate between the two above asymptotic behaviours are found numerically.
In numerical treatment,  we set $\mu=1$, $4\pi G_0=1$,
by using a scaled radial coordinate $r\to r \mu$ (together with $w\to w/ \mu$)
and scaled potentials
$H_1\to H_1 \sqrt{4\pi G_0}$, $V\to V \sqrt{4\pi G_0}$. 
Restricting to the fundamental set of solutions, we obtain the $(w,M)$-diagram, 
exhibited in Fig.~\ref{fig1}.
This spiral starts from $M = 0$ for $w/\mu = 1$, in which limit the Proca
field becomes very diluted and the solution trivializes.
At some intermediate frequency $w_{m}/\mu \simeq 0.875$, a maximal ADM
mass $M_{max} \simeq 1.058$ is attained. 
There is also a minimal allowed frequency
$w_{min} \simeq 0.814$ where $M\sim 0.82$.

\section{Code assessment}
\label{appendix}

\begin{figure}[t!]
\centering
\includegraphics[height=2.45in]{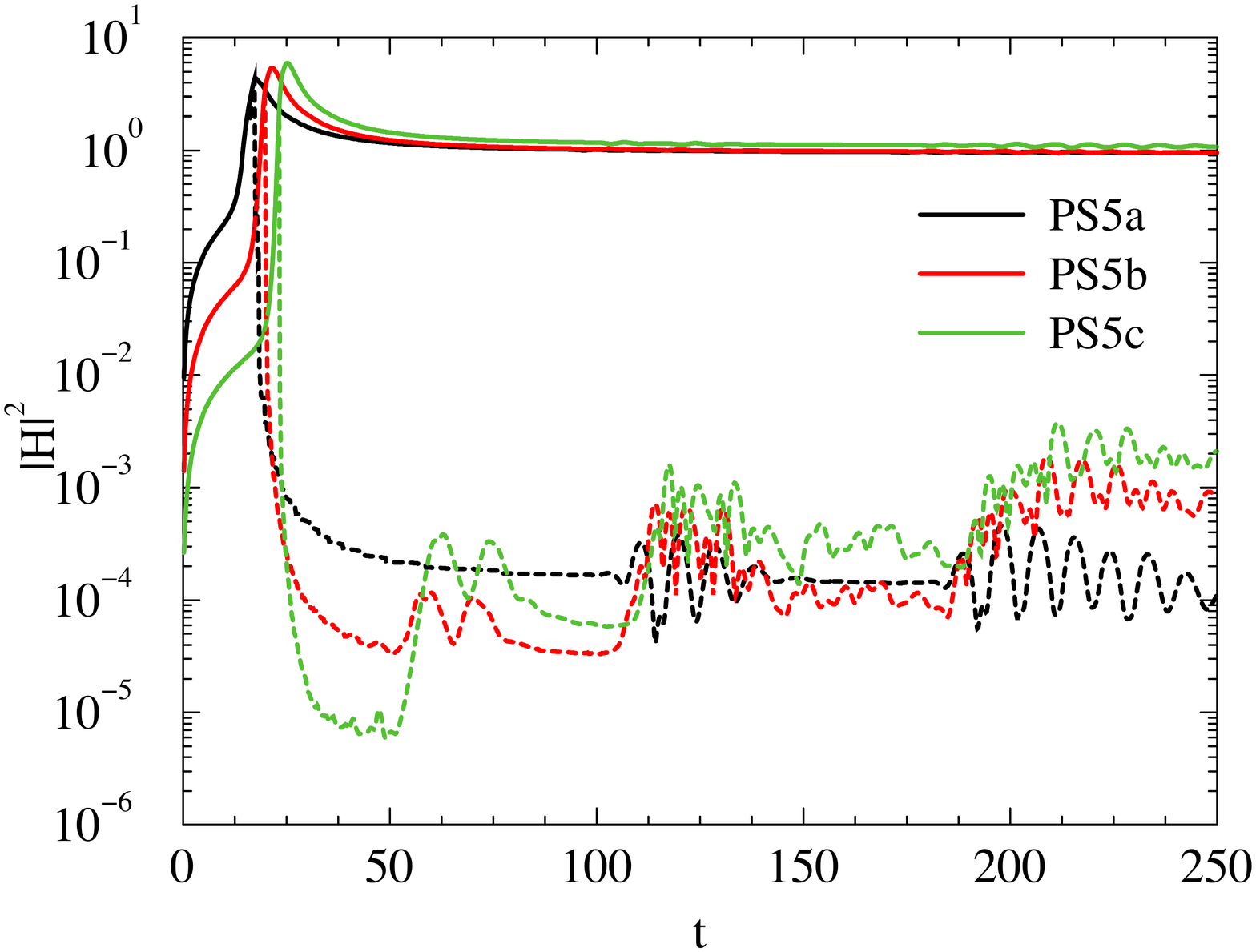}\\
\includegraphics[height=2.45in]{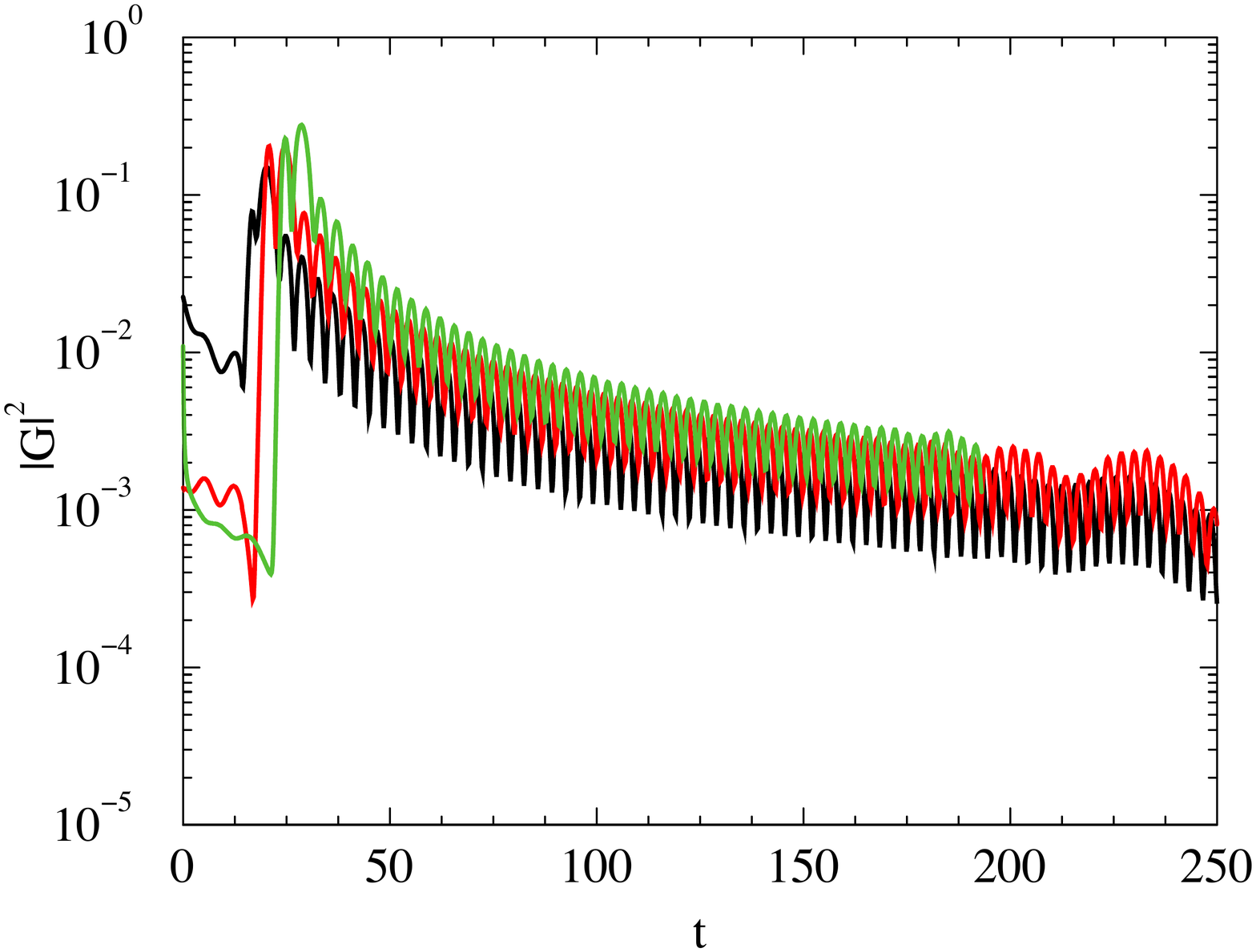}
\caption{Convergence analysis for model PS5 employing three different resolutions: $\Delta r=0.0125$, green curves, $\Delta r=0.00625$, red curves, and $\Delta r=0.003125$, black curves. Top panel: L2 norm of the Hamiltonian constraint computed either at all radial points (solid lines) or outside of the AH (dashed lines). Bottom panel: L2 norm of the Gauss constraint computed at all radial points. }
\label{fig6}
\end{figure}

Our numerical code was originally developed by \cite{baumgarte2013numerical}. The evolution equations are solved using techniques we have extensively tested and employed in previous 
works before~\cite{Montero:2012yr,Sanchis-Gual:2015bh,Sanchis-Gual:2015sms}. For the current work, we upgraded the code to account for the Proca field. This Appendix hence reports a succinct assessment of the upgraded code. We discuss in particular its convergence properties and the violations of the constraint equations, both for the Einstein and the Proca fields.


Fig.~\ref{fig6} shows the convergence properties of the code. This figure displays the time evolution of the rescaled L2 norm of the Hamiltonian constraint (top panel) and of the Gauss constraint (bottom panel) for the unstable model 5. The L2 norm of a constraint $C$ is given by
\begin{eqnarray}
|C|^{2}=\sqrt{\frac{\sum^{N}_{i=1}\,C_{i}^{2}}{N}}\,,
\label{L2norm}
\end{eqnarray}
where $N$ indicates the number of radial grid points. The results have been obtained at three different radial resolutions, as indicated in the figure caption, and the green and red curves have been multiplied by the appropriate numerical factors -- 4 and 16, respectively -- corresponding to second-order convergence (thus the three curves should overlap). The constraints are computed in the entire computational domain, hence the black hole puncture of this unstable model is also included. Black hole formation coincides with the rapid initial growth of the constraints. The violation of the Hamiltonian constraint levels off shortly thereafter and does not subsequently grow. Correspondingly, the violation of the Gauss constraint has decreased by about 2 orders of magnitude by the end of the simulation with respect to the value attained at the moment the black hole forms. In addition, the top panel of Fig.~\ref{fig6} also shows the Hamiltonian constraint when computed outside of the AH (as customary in numerical relativity studies, see e.g.~\cite{Alic:2013,Reisswig:2013}), which is indicated by the dashed lines. As expected, when the causally disconnected AH interior is excluded in the L2 norm computation, the violations are reduced by several orders of magnitude. All in all, Fig.~\ref{fig6} shows that the PIRK time-evolution scheme of our numerical code has an order of convergence close to 2.

\begin{figure}[t!]
\centering
\includegraphics[height=2.45in]{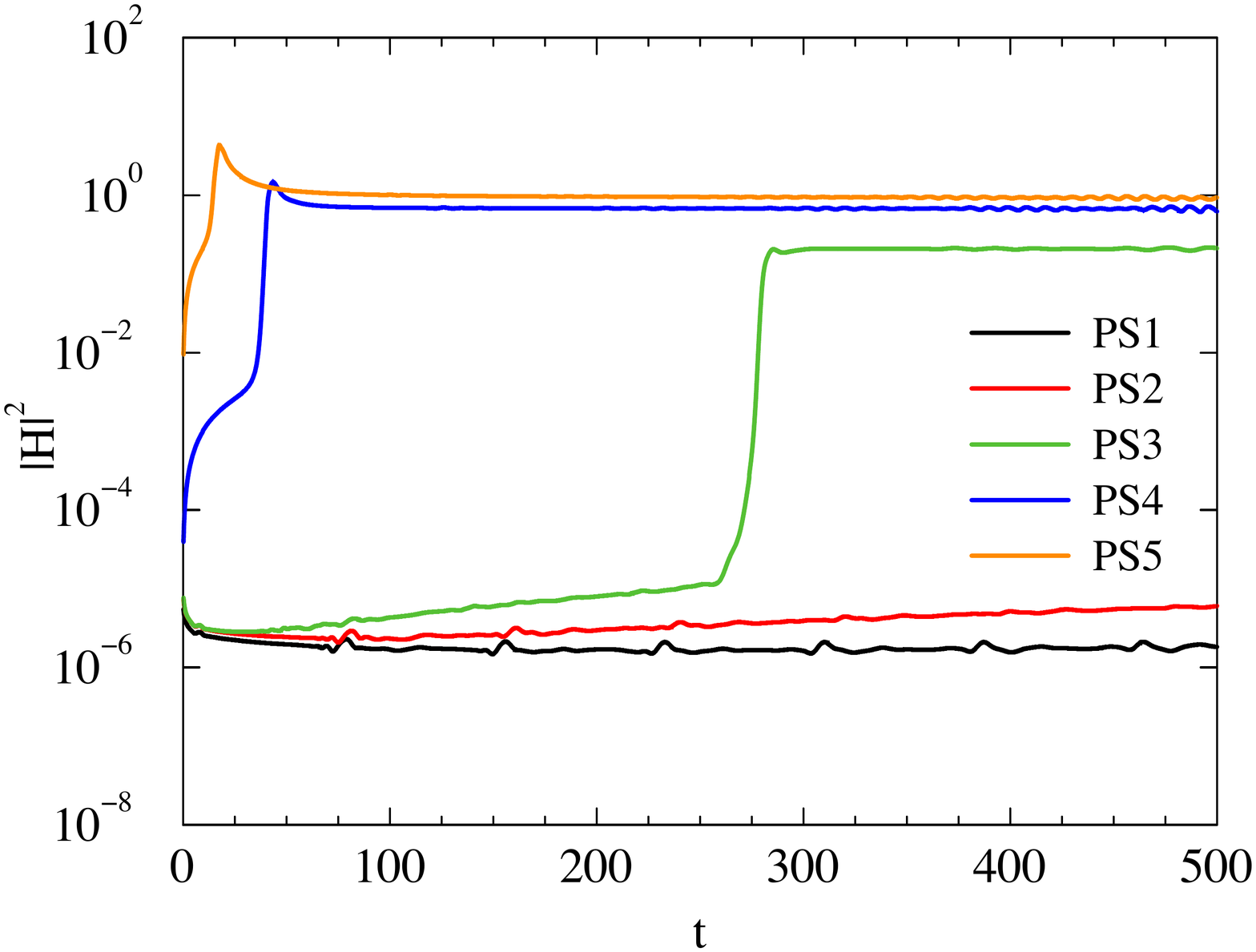}\\
\includegraphics[height=2.45in]{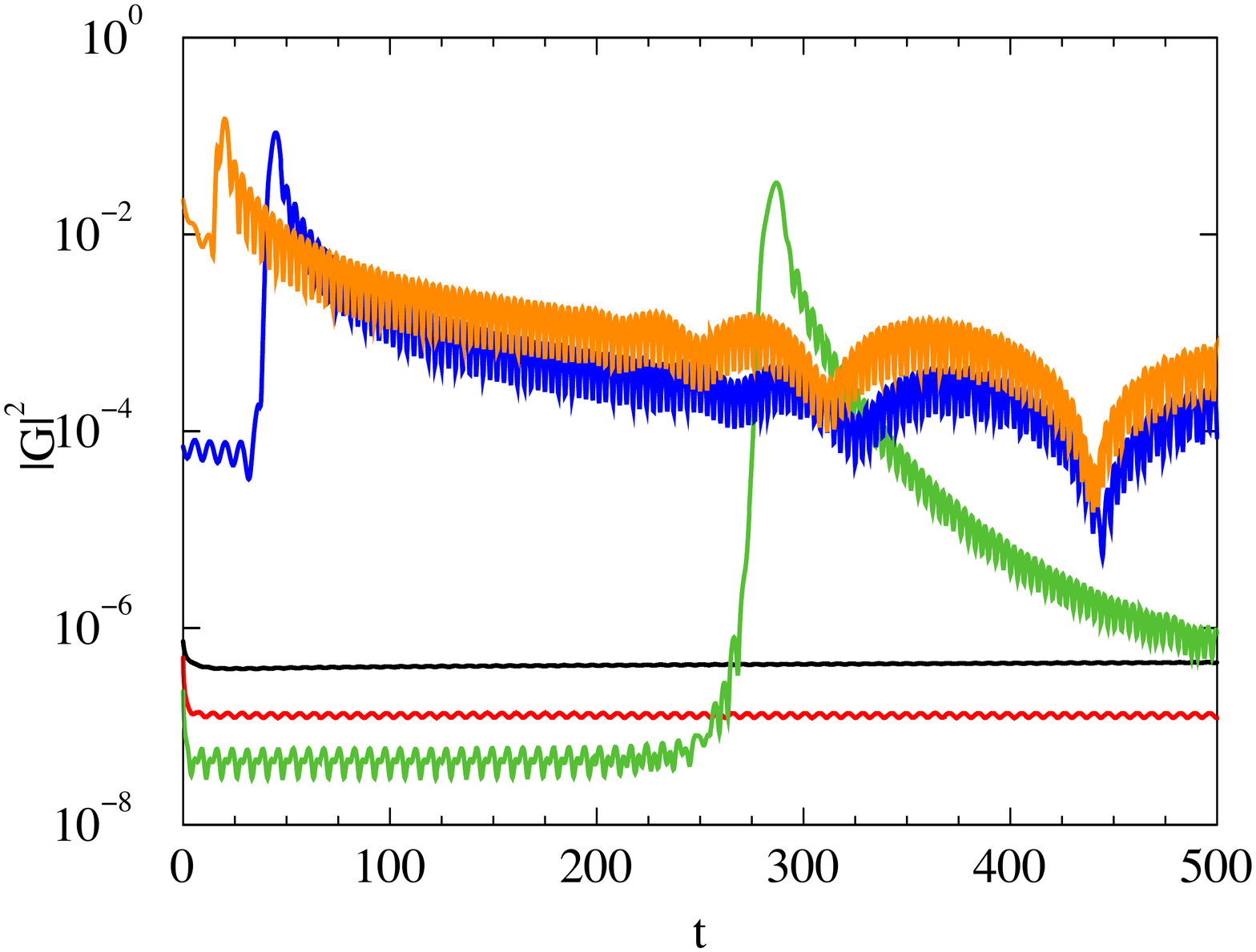}
\caption{Time evolution of the L2 norm of the constraints for all Proca star models, using our canonical radial resolution of $\Delta r=0.0125$. Top panel: Hamiltonian constraint. Bottom panel: Gauss constraint.}
\label{fig5}
\end{figure}


Finally, Fig.~\ref{fig5} shows the time evolution of the L2 norm of the Hamiltonian constraint and of the Gauss constraint for all of our Proca star models. Both quantities are computed using all radial grid points, hence including the AH of the resulting black holes for models 3, 4, and 5. The time evolution of the constraints for these models is characterized by the sudden growth at black hole formation, after which the Hamiltonian constraint attains a constant value and the Gauss constraint significantly drops. For the collapsing models, the higher the value of $\Phi_c$ the larger the violation of the constraints. On the other hand, for stable models 1 and 2, the violation of the constraints hardly grows in time and it stays at maximum values that are several orders of magnitude smaller than that of unstable models.

It should be noted, however, that the specific maximum violation the constraints attain is somewhat meaningless, as this number may be arbitrarily reduced even at the same grid resolution. This can be achieved by taking a larger number of grid points and placing the outer boundary at the appropriate (further out) location, as follows from the definition of the L2 norm in Eq.~(\ref{L2norm}). Therefore, what is important when assessing the numerical code is to show that the constraints are bounded in time {\it and} that the numerical code is convergent.

\end{appendix}


\bibliography{num-rel}

\end{document}